\documentstyle[epsfig,aps,prd,twocolumn]{revtex}

\def\etal{{\it et al.}}

\def\etc{{\it etc.}}
\def\ie{{\it i.e.}}

\def\lsim{\mathrel {\vcenter {\baselineskip 0pt \kern 0pt
    \hbox{$<$} \kern 0pt \hbox{$\sim$} }}}
    \def\gsim{\mathrel {\vcenter {\baselineskip 0pt \kern 0pt
    \hbox{$>$} \kern 0pt \hbox{$\sim$} }}}

\def\dddot#1{{\buildrel {. \kern-.05em . \kern-.05em .} \over {#1}}}

\begin{document}

\title{Limits on the Applicability of Classical Electromagnetic Fields as 
Inferred from the Radiation Reaction}

\author{Kirk T.~McDonald}
\address{Joseph Henry Laboratories, Princeton University, Princeton, NJ 08544}

\date{Jan.~29, 1998} 

\maketitle

\begin{abstract}
Can the wavelength of a classical electromagnetic field be arbitrarily small, or
its electric field strength be arbitrarily large?  If we require that the
radiation-reaction force 
on a charged particle in response to an applied
field be smaller than the Lorentz force we find limits on the classical
electromagnetic field that herald the need for a better theory, \ie, one
in better accord with experiment.  The classical
limitations find ready interpretation in quantum electrodynamics.  The
examples of Compton scattering and the QED critical field strength are
discussed.  It is still open to conjecture whether the present theory of QED is 
valid at field strengths beyond the critical field revealed by a semiclassical 
argument.
\end{abstract}

\section{Introduction}

The ultimate test of the applicability of a physical theory is the accuracy
with which it describes natural phenomena.  Yet on occasion the difficulty
of a theory in dealing with a ``thought experiment'' provides a clue as to
limitations of that theory.

It has long since been recognized that classical electrodynamics has been 
surplanted by quantum electrodynamics in some respects.  But one doubts
that quantum electrodynamics, or even its generalization, the Standard Model
of elementary particles, is valid in all domains.  To aid in the search
for new physics, it is helpful to review the warning signs of the
past transitions from one theoretical description to another.

The debates as to the meaning of the
classical radiation reaction for pointlike particles provide examples of 
such warning signs.  One case is the ``4/3 problem'' of electromagnetic
mass, where covariance does not imply uniqueness \cite{Schwinger83}.  Such
difficulties have often been interpreted as suggesting that classical
electrodynamics cannot be a complete description of matter on the scale of
the classical electron radius, $r_0 = e^2/mc^2$ in Gaussian units.

It seems less appreciated that the part of the classical radiation reaction
that is independent of particle size provides clues as to the limits of
applicability of classical electromagnetic fields.  
For example, a recent article
\cite{Rohrlich97} ends with the sentence, ``Only when all 
distances involved are in the classical domain is classical dynamics acceptable
for electrons''.  While this condition is necessary, it is not sufficient.
For a classical description to be accurate,
an electron can only be subject to fields that are not too strong.
This paper seeks to illustrate what ``not too strong'' means.

Considerations of strong fields have been very influential in the development
of other modern theories besides quantum electrodynamics.  In classical
gravity, \ie, general relativity, the strong-field problem is identified
with black holes.  One of the best known intersections between gravity and
quantum electrodynamics is the Hawking radiation of black holes.
In the case of the strong (nuclear) interaction, the fields associated with
nuclear matter all appear to be strong, and weak fields are
thought to exist only in the high-energy limit (asymptotic freedom).  
Such considerations led to
the introduction of non-Abelian gauge theories.  These constructs, when applied 
to the weak interaction, led to the concept of a background (Higgs) field 
that is strong in the sense of having a
large vacuum expectation value, which in turn has the effect of generating the 
masses of the
$W$ and $Z$ gauge bosons.  Most recently, considerations of the strong-field
(strong-coupling) 
limit of string theories have led to the notion of ``duality'', \ie, the
various string theories of the 1980's are actually different weak-field limits
of a single strong-field theory.  These string theories are noteworthy for
suggesting that particles
are to be considered as excitations of small, but extended quantum strings,
thereby avoiding the infinite self energies that have appeared in
 theories of point particles since J.J.~Thomson introduced the concept of
electromagnetic mass in 1881 \cite{Thomson81}.

The main argument concerning classical electromagnetic fields is given in 
sec.~2, and is brief.    This argument could have been given around 1900
by Lorentz \cite{Lorentz92,Lorentz03,Lorentz06} or by Planck \cite{Planck97}.  
who made remarks of a related nature.
But the argument seems to have been first made in 1935 by Oppenheimer 
\cite{Oppenheimer}, and more explicitly by Landau and Lifshitz \cite{Landau}.
Additional historical commentary is given in sec.~1.1.  Sections 2.1-2.5 
comment on various aspects of the main argument, still from a classical
perspective.  A quantum view in introduced in sec.~3, and the important
examples of Compton scattering and the QED critical field are discussed
in secs.~3.1-2.  The paper concludes in sec.~4 with remarks on the
role of strong fields on the development of  quantum electrodynamics, 
and presents two examples (secs.~4.1-2) of speculative features of strong-field 
QED and one of very short distance QED (sec.~4.3).

\subsection{Historical Introduction}

The relation between Newton's third law and electromagnetism has been of
concern at least since the investigations of Amp\`ere, who insisted that
the force of one current element on another be along their line of centers.
See Part IV, Chap.~II, especially sec.~527, of Maxwell's {\sl Treatise} for a
review \cite{Maxwell}.  However, the presently used differential version of
the Biot-Savart law does not satisfy Newton's third law for pairs of current
elements unless they are parallel.

Perhaps discomfort with this fact contributed to the delay in acceptance of 
the concept of isolated electrical charges, in contrast to complete loops of
current, until the late nineteenth century.

A way out of this dilemma became possible after 1884 when Poynting 
\cite{Poynting} and Heaviside \cite{Heaviside}
argued that electromagnetic fields (in suitable configurations) can be thought 
of as transmitting energy.  The transmission of energy was then extended 
by Thomson \cite{Thomson}, Poincar\'e \cite{Poincare00} and
Abraham \cite{Abraham02} to include transmission (and storage) of 
momentum by an electromagnetic field.

That a moving charge interacting with thermal radiation should feel a
radiation pressure was anticipated by Stewart in 1871 \cite{Stewart}, who
inferred that both the energy and the momentum of the charge would be affected.

In 1873, Maxwell discussed the pressure of light on conducting media at rest,
and on ``the medium in which waves are propagated''
(\cite{Maxwell}, secs.~792-793). 
In the former case, the radiation of a reflected wave by a (perfectly
conducting) medium in response
to an incident wave results in momentum, but not energy, being transferred to
the medium.  The energy for the reflected wave comes from the incident wave.

The present formulation of the radiation reaction is due
to Lorentz' investigations of the self force of an extended
electron, beginning in 1892 \cite {Lorentz92} and continuing through 1903
\cite{Lorentz03}.  The example of dipole radiation of a single charge 
contrasts strikingly with Maxwell's discussion of reaction forces during
specular reflection.  There is no net momentum
radiated by an oscillating charge with zero average velocity, but energy
is radiated.  The external force alone can not account for the energy 
balance.  An additional force is needed, and was identified by Lorentz as the
net electromagnetic force of one part of an extended, accelerated  charge 
distribution on another.  See eq.~(\ref{eq1}) below.

In 1897, Planck \cite{Planck97,Planck06} applied the radiation reaction force
of Lorentz to a model of charged oscillators and noted the existence of
what are sometimes called ``runaway'' solutions, which he dismissed as having
no physical meaning (keine physikalishe Bedeutung).

The basic concepts of the radiation reaction were brought 
essentially to their final form by Abraham \cite{Abraham03,Abraham04}, 
who emphasized the balance of both energy and momentum in the motion of 
extended electrons moving with arbitrary velocity.

Important contributions to the subject in the early twentieth century
include those of Sommerfeld \cite{Sommerfeld},
Poincar\'e \cite {Poincare}, Larmor \cite{Larmor}, Lindemann \cite{Lindemann},
 Von Laue \cite{vonLaue09}, Born \cite{Born},
Schott \cite{Schott,Schott15,Schott27,Schott33}, Page \cite{Page18},
Nordstr\"om \cite{Nordstrom}, Milner \cite{Milner21}, Fermi \cite{Fermi},
Wenzel \cite{Wenzel}, Wesel \cite{Wesel} and Wilson \cite{Wilson}.
The main theme of these works
was, however, models of classical charges and the related topic of
electromagnetic mass.

The struggle to understand the physics of atoms led to diminished attention
to classical models of charged particles in favor of quantum mechanics and 
quantum
electrodynamics (QED).  In 1935, there was apparent disagreement between
QED and reported observations at energy scales of 10-100 MeV. 
Oppenheimer \cite{Oppenheimer} then
conjectured whether QED might fail at high energies and, in partial support
of his view, invoked a classical argument concerning difficulties of
interpretation of the radiation reaction at short distances.  The present
article illustrates an aspect of Oppenheimer's argument that was developed
further by Landau \cite{Landau}.

Another response to Oppenheimer's conjecture was by
Dirac \cite{Dirac} in 1938, when he deduced a covariant expression for
the radiation reaction force (previously given by Abraham, Lorentz and von Laue
in noncovariant notation)
by an argument not based on a model of an extended electron.
Dirac also gave considerable discussion of the paradoxes of runaway
solutions and pre-acceleration.  This work of Dirac, and most subsequent
work on the classical radiation reaction, emphasized the internal consistency
of classical electromagnetism as a mathematical theory, rather than as a
description of nature.  But, as has been remarked by Schott \cite{Schott}, 
``there is considerable danger, in a purely mathematical investigation, of
losing touch with reality''.  Quantum mechanics had triumphed.

Research articles on the classical
radiation reaction are still being produced; see, for example, 
Refs.~\cite{Pryce}-\cite{Hartemann96}.  Sarachik and Schappert 
(\cite{Sarachik}, sec.~IIID) present a brief version of the argument given
below in sec.~2.

Reviews of the subject include Refs.~\cite{Abraham05}-\cite{Jimenez}.
Most noteworthy in relation to the present article are the reviews by
Lorentz \cite{Lorentz06}, Erber \cite{Erber} and Klepikov \cite{Klepikov}, 
which are the only ones
that indicate an awareness of the problem of strong fields.  The texts of
Landau and Lifshitz \cite{Landau}, Jackson \cite{Jackson} and
Milonni \cite{Milonni} briefly mention
that issue.

The radiation reaction has
been a frequent topic of articles in the American Journal of Physics, including 
Refs.~\cite{Rohrlich97} and \cite{Page45}-\cite{Stump}.  
The article of Page and Adams
\cite{Page45} is noteworthy for illustrating how the concept of electromagnetic
field momentum restores the full validity of Newton's third law in an
interesting example of the interaction of a pair of moving charges.

\section{A General Result for the Radiation Reaction}

Consider an electron of charge $e$ and mass $m$ moving in electric and
magnetic fields {\bf E} and {\bf B}.  The mass $m$ is the 
``effective mass'' in the language of Lorentz \cite{Lorentz06}, now called the 
``renormalized'' mass, for which the divergent electromagnetic self energy of a 
small electron is cancelled in a manner beyond the scope of this article.
Then the remaining leading effect of the radiation
reaction is the ``radiation resistance'' which is independent of hypotheses 
as to the structure of the electron.  
Our argument emphasizes the effect of radiation resistance, since any deductions
about properties of electromagnetic fields will then be as free as possible
from controversy as to the nature of matter.

The (nonrelativistic) equation of motion 
including radiation resistance is (in Gaussian units)
\begin{equation}
m\dot{\bf v} = {\bf F}_{\rm ext} + {\bf F}_{\rm resist},
\label{eq1}
\end{equation}
where 
\begin{equation}
{\bf F}_{\rm ext}  =  e{\bf E} + e{{\bf v} \over c} \times {\bf B} 
\label{eq1a}
\end{equation}
is the Lorentz force on the electron due to the external field,
\begin{equation}
{\bf F}_{\rm resist} = {2e^2 \over 3c^3}\ddot{{\bf v}} + {\cal O}({\bf v}/c)
\label{eq1b}
\end{equation}
is the force of radiation resistance, {\bf v} is the velocity of the
electron, $c$ is the speed of light and the dot indicates differentiation
with respect to time.
 Equation~(\ref{eq1b}) is the form of the radiation reaction given 
in the  original derivations of
Lorentz \cite{Lorentz92} and Planck \cite{Planck97,Planck06},
which is sufficient for the main argument of this
paper.   Some discussion of the larger context of the classical radiation 
reaction is given in secs.~2.1-5.

If the second time derivative of the velocity is small we estimate it 
by taking the derivative of (\ref{eq1}):
\begin{equation}
\ddot{\bf v} \approx {e \dot{\bf E} \over m} 
     + {e \over m}{\dot{\bf v} \over c} \times {\bf B} 
     + {e \over m}{{\bf v} \over c} \times \dot{\bf B}.
\label{eq2}
\end{equation}
We further suppose that the velocity 
is small (without loss of generality according to the principle of relativity;
see sec.~2.4 for a  relativistic discussion).
so it suffices to approximate $\dot{\bf v}$ as $e{\bf E}/m$ in (\ref{eq2}).
Hence,
\begin{equation}
\ddot{\bf v} \approx {e \dot{\bf E} \over m} + {e^2 \over m^2c}{\bf E} 
\times {\bf B}.
\label{eq3}
\end{equation}
The radiation resistance is now
\begin{equation}
{\bf F}_{\rm resist} \approx {2e^2 \over 3c^3} \left( {e \dot{\bf E} \over m} 
     + {e^2 \over m^2c}{\bf E} \times {\bf B} \right).
\label{eq4}
\end{equation}
The first term in (\ref{eq4}) contributes only for time-varying fields, 
which I take to have frequency $\omega$ and reduced wavelength $\lambdabar$; 
hence, $\dot{\bf E} \propto \omega {\bf E}$.  The
second term contributes only when ${\bf E} \times {\bf B} \neq 0$, which is
most likely to be in a wave (with $E =B$) if the fields are large.  So, for an
electron in an external wave field, the magnitude of the radiation-resistance
force is
\begin{eqnarray}
F_{\rm resist} & \approx & {2 \over 3} eE \sqrt{ \left( {e^2 \over mc^2} 
{\omega \over c} \right)^2 + \left( {e^3 E \over m^2c^4} \right)^2}
\nonumber \\
& \approx & F_{\rm ext} \sqrt{ \left( {r_0 \over \lambdabar} \right)^2 
            + \left( {E \over  e/r_0^2} \right)^2 },
\label{eq5}
\end{eqnarray}
where $r_0 = e^2/mc^2 = 2.8 \times 10^{-13}$ cm is the classical electron
radius.

Equation (\ref{eq5}) makes physical sense only 
when the radiation reaction 
force is smaller than the external force.  Here we don't explore whether the 
length $r_0$
describes a physical electron; we simply consider it to be a length that
arises from the charge and mass of an electron.  Rather, we concentrate on
the implication of eq.~(\ref{eq5}) for the electromagnetic field.  Then we
infer that a classical description becomes implausible for fields whose
wavelength is small compared to length $r_0$, or whose strength is large 
compared to $e/r_0^2$.

\subsection{Commentary}

The argument related to eq.~(\ref{eq5}) is that
there are classical electromagnetic fields that lead to physically
implausible behavior when radiation-reaction effects are included.  This does
not necessarily imply any mathematical inconsistency in the theory.  Indeed,
various authors have displayed solutions for electron motion coupled to an
oscillator of very high natural frequency \cite{Loinger,Plass}.  Such
solutions are well-defined mathematically but appear ``physically implausible''.
Of course, the mathematics might be correct in predicting the physical
behavior in an unfamiliar situation.  So it becomes a matter of experiment to
decide whether the characterization ``implausible'' corresponds to physical 
reality or not.  The experiments that produce the most influential results
are typically those that reveal new phenomena in realms where prevailing
theories are ``implausible''.

 Thus far, there is no evidence for the behavior predicted by the
classical equations for electrons interacting with waves of frequencies
greater than $c/r_0$.  Rather, quantum mechanics is needed for
a good description of the phenomena observed in that case, Compton
scattering being an early example (sec.~3.1).  
Laboratory studies of strong-field electrodynamics  have been undertaking only 
recently (sec.~3.2), and deal primarily with effects not anticipated in a
classical description.

The argument of sec.~2 can also be considered as a model-independent version of
a restriction that Lorentz placed on his derivation of 
eqs.~(\ref{eq1}-\ref{eq1b}) (\cite{Lorentz06}, sec.~37, eq.~(73)).
Namely, the derivation makes physical sense only if 
\begin{equation}
{l \over ct} \ll 1,
\label{eq90}
\end{equation}
where $l$ is a characteristic length of the problem, and $t$ is 
a characteristic time ``during which the state of motion is sensibly altered''.

Lorentz would certainly have considered the classical electron radius, $r_0$, as
an example of a relevant characteristic length.
Hence, for an electron in an electromagnetic wave of (reduced) wavelength 
$\lambdabar$, the characteristic time of the resulting motion is 
$\lambdabar/c$, and Lorentz' condition (\ref{eq90}) becomes
\begin{equation}
{r_0 \over \lambdabar} \ll 1,
\label{eq91}
\end{equation}
A close variant of the above argument was also given by Planck \cite{Planck97}.

In case of a strong field with a long (possibly infinite) wavelength, 
Lorentz' condition (\ref{eq90}) can be interpreted as
requiring the change in the electron's velocity to be small compared to
the speed of light during the time it takes light to travel one
classical electron radius.  That is, we require
\begin{equation}
\Delta v = a \Delta t = {e E  \over m} {r_0 \over c}  \ll c,
\label{eq92}
\end{equation}
and hence,
\begin{equation}
E \ll {mc^2 \over er_0} = {e \over r_0^2}.
\label{eq93}
\end{equation}
Thus, we arrive by another (although closely related) path to the conclusion
drawn previously from eq.~(\ref{eq5}).   Perhaps because the limiting field
strength implied by (\ref{eq93}) is extraordinarily large by practical
standards, neither Lorentz nor Planck mentioned it explicitly.

In the first sentence of his 1938 article, Dirac \cite{Dirac} stated that
``the Lorentz model of the electron...has proved very valuable...in a 
certain domain of problems, in which the electromagnetic field does not
vary too rapidly and the accelerations of the electrons are not too great''.
However, he does not elaborate on the meaning of ``not too great''.

Dirac's derivation of the radiation-reaction 4-force was not based on a model
of an extended electron, and so the derivation was not subject to Lorentz'
restriction (\ref{eq90}).  But as a co-inventor of quantum mechanics,
Dirac cannot have expected his classical results to have unrestricted
validity in the physical world.

In the decade after Dirac's 1938 paper, a few works
\cite{Wesel,Pomeranchuk,Bopp,Stueckelberg,Feynman48,Haag}
 appeared that commented
on the concept of a limiting field strength, typically in classical
discussions of electron-positron pair creation.  In sec.~3.2 we return to the 
issue of pair creation, but in a quantum context.

After the discovery of pulsars in 1967 there was a burst of interest in the
behavior of electrons in very strong magnetic fields.  Several papers
appeared in which classical electrodynamics was applied
\cite{Shen70a,Sengupta70,Shen70b,Sengupta72,Jaffe,Shen72b,Herrera73,Shen78},
often with the
intent of clarifying the boundary between the classical and quantum domains.
For very large fields, classical solutions to the motion were obtained in 
which the electron has a damping time constant that is small compared to
the period of cyclotron motion.  Whether or not such highly damped solutions 
are ``implausible'', they are outside ordinary experience.  Again, one must
perform experiments to decide whether the classical theory is
valid in this domain.  
If such experiments had been possible prior to the development of
quantum mechanics, they would have revealed deviations from the classical
theory that would have encouraged development of a new theory.  
Arguments such as those leading to eqs.~(\ref{eq5}), 
(\ref{eq91}) and (\ref{eq93})
would have motivated the experiments.

\subsection{Another Strong-Field Regime}

Are there any other domains in which classical electrodynamics might be
called into question?

Another interpretation of Lorentz' criterion (\ref{eq90}) is that the
amplitude of the oscillatory motion of an electron in a wave of frequency 
$\omega$ should be small compared to the wavelength.  As is well known
(see prob.~2, sec.~47 of Ref.~\cite{Landau}), this leads to the condition
that the dimensionless, Lorentz-invariant quantity,
\begin{equation}
\eta = {e E_{\rm rms} \over m \omega c},
\label{eq94}
\end{equation}
should be small compared to one.  Parameter $\eta$ can exceed unity for
waves of very low field strength if the frequency is low enough.  
An interesting result is that the electron can be said to have an
effective mass,
\begin{equation}
m_{\rm eff} = m \sqrt{1 + \eta^2},
\label{eq95}
\end{equation}
when inside a wave field \cite{Kibble}.  An electron in a spatially varying
wave experiences a force ${\bf F} = -\nabla m_{\rm eff} c^2$ which is often 
called the ``ponderomotive force'', but which can be regarded as a kind of
radiation pressure for a case where the ``reflected'' wave cannot be 
distinguished from the incident wave, and hence as a kind of
radiation reaction force in its broadest meaning.

Debates continue regarding energy-transfer mechanisms
between electrons and strong classical waves (as represented by a laser beam
with $\eta \gsim 1$).  To what extent can net energy be exchanged between a
free electron and a laser pulse in vacuum?  Does a classical discussion suffice?
Our understanding suggests that quantum aspects
should be unimportant even for $\eta \gg 1$ so long as condition
(\ref{eq93}) is satisfied,
but full understanding has been elusive.
Detailed discussion of this matter is deferred to a future article.

\subsection{Utility of the Classical Radiation Reaction}

Besides provoking extensive discussion on the validity of classical
electrodynamics, the radiation reaction has enjoyed some well-known
success in classical phenomenology.  In particular, the topics of
linewidth of radiation by a classical oscillator and resonance width in
scattering of waves off such an oscillator show how partial understanding
of atomic systems can be obtained in a classical context.  Also, the
radiation reaction is very important in antenna engineering where the
power source must provide for the energy (and momentum, if any) radiated as 
well as that consumed in Joule losses.  It is worth noting
that these successes hold where the electron is part of an extended system.

In contrast, the radiation reaction has been almost completely negligible in
descriptions of the radiation of free electrons for practical parameters in the 
classical domain (\ie, outside the domain of quantum mechanics).  That this
might be the case is the main argument of sec.~2.  Section 3 discusses
effects of the radiation reaction in the quantum domain.

\subsection{Relativistic Radiation Reaction}

For purposes of additional commentary, it is useful to record 
relativistic expressions for the radiation reaction.

The relativistic version of (\ref{eq1}) in 4-vector notation is
\begin{equation}
mc^2 {du^\mu \over ds} = F^\mu_{\rm ext} + F^\mu_{\rm resist},
\label{eq110}
\end{equation}
with external 4-force $F^\mu_{\rm ext} = \gamma({\bf F}_{\rm ext} \cdot 
{\bf v}/c, {\bf F}_{\rm ext})$, and radiation-reaction 4-force given by
\begin{equation}
F^\mu_{\rm resist} = {2e^2 \over 3} {d^2 u^\mu \over ds^2} - {R u^\mu \over c}, 
\label{eq111}
\end{equation}
where
\begin{equation}
R = - {2e^2 c \over 3} {du_\nu \over ds} {du^\nu \over ds}
= {2e^2 \gamma^6 \over 3 c^3} \left[ \dot{\bf v}^2 - {({\bf v} \times
\dot{\bf v} )^2 \over c^2} \right] \geq 0
\label{eq112}
\end{equation}
is the invariant rate of radiation of energy of an accelerated charge,
$u^\mu = \gamma(1,{\bf v}/c)$ is the 4-velocity, 
$\gamma = 1/\sqrt{1 - v^2/c^2}$,  $ds = cd\tau$ is the 
invariant interval and the metric is $(1,-1,-1,-1)$.   

The time component of eq.~(\ref{eq110}) can be written
\begin{equation}
{d\gamma m c^2 \over dt} = {\bf F}_{\rm ext} \cdot {\bf v} +
{d \over dt} \left( {2 e^2 \gamma^4 {\bf v} \cdot \dot{\bf v} \over 3 c^2}
\right) - R,
\label{eq113}
\end{equation}
and the space components as
\begin{eqnarray}
& & {d\gamma m {\bf v} \over dt} = {\bf F}_{\rm ext} 
\label{eq114} \\
& + & {2 e^2 \gamma^2  \over 3 c^3} \left[ \ddot {\bf v} 
+ {3 \gamma^2 \over c^2} ({\bf v} \cdot \dot{\bf v}) \dot{\bf v} 
+ { \gamma^2 \over c^2} ({\bf v} \cdot \ddot{\bf v}) {\bf v} 
+ {3 \gamma^4 \over c^4} ({\bf v} \cdot \dot{\bf v})^2 {\bf v} 
\right]. 
\nonumber
\end{eqnarray}
Keeping terms only to first order in velocity, eqs.~(\ref{eq113}-\ref{eq114})
become
\begin{equation}
{d m v^2/2 \over dt} 
= {\bf F}_{\rm ext} \cdot {\bf v} +
 {2 e^2 {\bf v} \cdot \ddot{\bf v} \over 3 c^3},
\label{eq115}
\end{equation}
and
\begin{equation}
{dm {\bf v} \over dt} = {\bf F}_{\rm ext} + {2 e^2 \ddot {\bf v} \over 3 c^3} 
+ {2 e^2 ({\bf v} \cdot \dot{\bf v}) \dot{\bf v} \over c^3}. 
\label{eq116}
\end{equation}

Equations (\ref{eq113}-\ref{eq114}) were first given by Abraham 
\cite{Abraham04}.  Von Laue \cite{vonLaue09} was the first to show that these
equations can be obtained by a Lorentz transformation of the
nonrelativistic results (\ref{eq115}-\ref{eq116}).  The covariant
notation of eqs.~(\ref{eq110}-\ref{eq112}) was first applied to the
radiation reaction by Dirac \cite{Dirac}.  An interesting discussion of the
development of eqs.~(\ref{eq113}-\ref{eq114}) has been given recently by
Yaghjian \cite{Yaghjian}.

\subsection{Terminology}

During a century of discussion of the radiation reaction a variety of
terminology has been employed.  
In this article I use the phrase ``radiation reaction'' to cover all aspects
of the physics of ``R\"uckwirkung der Strahlung'' as introduced by
Lorentz and Abraham.  This usage contrasts with
a proposed narrow interpretation discussed at the end of this section.

``\AE thereal friction'' was the first description by Stewart \cite{Stewart} 
in 1871, which he used in only a qualitative manner.  

In 1873, Maxwell wrote on the ``pressure exerted by light'' in secs.~792-793
of his {\sl Treatise} \cite{Maxwell}.

Lorentz used the French word ``r\'esistance'' in describing
eq.~(\ref{eq1b}) when he presented it in 1892, and used the
English equivalent ``resistance'' in his 1906 Columbia lectures
\cite{Lorentz06}.  

Planck \cite{Planck97,Planck06} 
also discussed eq.~(\ref{eq1b}), which he
described as ``D\"ampfung'' (damping) and ``D\"ampfung durch Strahlung''
(literally, ``damping by radiation'' but translated more smoothly as
``radiation damping'').  The term ``Strahlungsd\"ampfung''
(radiation damping) does not, however, appear in the German literature 
until 1933 \cite{Wenzel}.

Around 1900, Larmor \cite{Larmor} used the terms ``frictional resistance'' and
``ray pressure'' to describe a result meant to quantify Stewart's insight,
but which analysis has not stood the test of time.

The massive analyses of Abraham were accompanied by 
the introduction of several new terms.  The title of Abraham's 1904 article
\cite{Abraham04} included the term ``Strahlungsdruck'' (radiation pressure).
This use of the phrase ``radiation pressure'' can, however, be confused with 
the simpler concept of
the pressure that results when a wave is reflected from a conducting surface
\cite{Maxwell}.
Perhaps for this reason, Abraham also introduced the phrase
``Reaktionskraft der Strahlung'',
which I translate as ``radiation reaction force''.   This appears to be the
origin of the phrase ``radiation reaction'', although in German that phrase
remained a qualifier to ``Kraft'' (force) for many years.  The variant
``Strahlungsreaktion'' (radiation reaction) appeared for the first time in
1933 \cite{Wesel}.

Lorentz' 1903 Encyklop\"adie article \cite{Lorentz03} introduced the topic
of the radiation reaction with the phrase ``R\"uckwirkung des \"Athers''
(back interaction of the \ae ther).
In his 1905 monograph \cite{Abraham05}, Abraham used the variant
``R\"uckwirkung der Strahlung'' (back interaction of radiation, which could
also be translated agreeably as ``radiation reaction'').

In England in 1908, the Adams Prize examiners chose the topic of the radiation
reaction, suggesting the cumbersome title ``The Radiation from Electric
Systems or Ions in Accelerated Motion and the Mechanical Reactions on their
Motion which arise from it''.  The winning essay by Schott \cite{Schott}
adopted much of this title, but in the text Schott refers to ``radiation
pressure'' and indicates that he follows Abraham in this.  
In his 1915 article, Schott \cite{Schott15} also used the phrase ``reaction due
to radiation'' and indicated that it was equivalent to his use of the
phrase ``radiation pressure''.

Schott also introduced other terms that seem less than ideal descriptions of
the phenomena associated with the radiation reaction.  His argument of
1912 \cite{Schott} is less crisp than one he gave in 1915 \cite{Schott15},
so I follow the latter here.  Schott considered the rate 
at which a radiating charge loses energy, and deduced eq.~(\ref{eq113}) in
essentially that form.
Schott noted that term $R$ is just the
rate of radiation of energy by an accelerated charge, which he described as
an ``irreversible'' process.  He then interpreted the term
\begin{equation}
Q = {2\gamma^4 e^2 {\bf v} \cdot \dot{\bf v} \over 3c^3},
\label{eq102}
\end{equation}
as an energy stored ``in the electron in virtue of its
acceleration'' and gave it the name ``acceleration energy''.  
Schott considered the term $\dot Q$ in eq.~(\ref{eq113}) 
to be a ``reversible'' loss of energy.

Insights related to the concept of the ``acceleration energy'' have been 
useful in resolving the paradox of whether a charge radiates if its
acceleration is uniform, \ie, if $\ddot{\bf v} = 0$.  In this case the
radiation reaction force (\ref{eq1b}) vanishes and many people have
argued that this means there is no radiation 
\cite{Born,Nordstrom,Pauli,Feynman}.
But as first argued by Schott \cite{Schott15}, in the case of uniform
acceleration ``the energy radiated by the electron is derived entirely from
its acceleration energy; there is as it were internal compensation amongst
the different parts of its radiation pressure, which causes its resultant
effect to vanish''.  This view is somewhat easier to follow if 
``acceleration energy'' means energy stored in the near and induction zones of 
the electromagnetic field \cite{Fulton,Thirring}.

Schott's use of the word ``irreversible'' to describe the process of
radiation seems inapt.  He may have meant that in a classical
universe containing only one electron and an external force field, 
the radiated energy can never return
to the electron.  But as noted by Planck \cite{Planck14},
``the fundamental equations of mechanics as well as those of electrodynamics
allow the direct reversal of every process as regards time''.   For example,
``if we now consider any radiation processes whatsoever, taking place in a 
perfect vacuum enclosed by reflecting walls, it is found that, since they are
completely determined by the principles of classical electrodynamics, there can
be in their case no question of irreversibility of any kind''.  However,
``an irreversible element is introduced by the addition of emitting and 
absorbing substance''.  Thus, consistent use of the word ``irreversible''
goes beyond classical electron theory.  These views of Planck were seconded by
Einstein \cite{Ritz} and elaborated upon in
the absorber theory of radiation of Wheeler and Feynman \cite{Wheeler}.  

As another counterexample to the view that radiation is irreversible, 
a theme of contemporary
accelerator physics is that every radiation process can be inverted to
produce energy gain, not loss.  Hence, there are now devices that accelerate
electrons based on inverse \v Cerenkov
radiation, inverse free-electron radiation, inverse Smith-Purcell radiation,
inverse transition radiation, \etc\ \
Uniform acceleration is the inverse of uniform deceleration, and
the inverse transformation is especially simple here: since 
${\bf F}_{\rm resist}$ vanishes, it suffices to reverse the sign
of the external force.  These inverse radiation processes will be the
subject of a future paper.

Schott's use of ``irreversible'' as applied to the term $-Ru^\mu/c$ of the
radiation reaction has not
been followed in the German literature.  See Ref.~\cite{Thirring} for an
interesting contrast.

The English phrase ``radiation reaction'' appears to have been first used
by Page in 1918 \cite{Page18}.

In his 1938 paper, Dirac \cite{Dirac} used the phrase ``the effect of radiation
damping on the motion of the electron''.  As a consequence, most subsequent
papers use ``radiation damping'' interchangeably with ``radiation reaction''
as a general description of the subject.  Thus, in German there appeared
the use of ``Strahlungsd\"amfung'' \cite{Wenzel} (already in 1933), in French,
``force de freinage'' \cite{Stueckelberg} (braking force, compare
``rayonnement freinage'' = Bremsstrahlung), and in Russian the equivalent of
``radiation damping'' must have been used as well \cite{Landau}.
Dirac seconded Schott's use of
the terms ``irreversible'' and ``acceleration energy'', and these become
fairly common in the English literature thereafter.  Indeed, ``acceleration
energy'' becomes ``Schott acceleration energy'', or just ``Schott energy''.

The terminology of Schott and Dirac was taken a step further by Rohrlich
in 1961 \cite{Rohrlich61} and 1965 \cite{Rohrlich65},
who proposed that only the second term in
the covariant expression (\ref{eq111}) is entitled to be called ``the
radiation reaction''.  The first term of (\ref{eq111}) is to be called
the ``Schott term''.  
A motivation for this terminology appears to be that in the
case of uniform acceleration, expression (\ref{eq111}) vanishes 
by virtue of cancellation of its two nonzero terms.  Then the broadly
defined ``radiation reaction'' (\ie, eq.~(\ref{eq111})) vanishes, 
but the radiation does not (although it takes considerable effort to 
demonstrate this \cite{Fulton}).  The
terminology of Rohrlich has the merit that the paradox ``how can there be
radiation if there is no radiation reaction'' is avoided
in this case since only
the (nonvanishing) term $-Ru^\mu/c$ is called the ``radiation reaction''.

However, this terminology is at odds with the origins of the subject,
which emphasize the low-velocity limit, eqs.~(\ref{eq115}-\ref{eq116}).  Here,
the radiated momentum enters only in terms of order $v^2/c^2$, so the
direct back reaction of the radiated momentum (\ie, $-Ru^\mu/c$) plays no role 
in the nonrelativistic limit.  Thus, according to Rorhlich's terminology there
is no ``radiation reaction'' in the nonrelativistic limit.

But the original, and continuing, purpose of the concepts of the radiation
reaction is to describe how a charge reacts to the radiation of energy
when it does not radiate net momentum.  To define the ``radiation reaction'' to
be zero in this circumstance is counterproductive.

It appears that the terminology of Rohrlich has been adopted only by three
subsequent workers \cite{Teitelboim,Sarachik,Pearle}.

\section{A Quantum Interpretation}

To go further, we pass beyond the realm of classical electromagnetism.
The remainder of this paper is not a direct consequence of that theory,
but considers how only a modest admixture of quantum concepts greatly
clarifies the hints deduced by classical argument.

A simple device is to multiply and divide eq.~(\ref{eq5}) by Planck's
constant $\hbar$, which was introduced by him \cite{Planck00}
shortly after his work on eq.~(\ref{eq1}) \cite{Planck97}.  Then we can write
\begin{eqnarray}
F_{\rm resist} & \approx & F_{\rm ext} \sqrt{ \left( {e^2 \over \hbar c}
{\hbar \over mc} {\omega \over c} \right)^2 
   + \left( {e^2 \over \hbar c} {e\hbar \over m^2c^3} E \right)^2 }
\nonumber \\
& \approx & \alpha F_{\rm ext} \sqrt{ \left( {\lambdabar_C  \over \lambdabar} 
           \right)^2 + \left( {E \over  E_{\rm crit}} \right)^2 },
\label{eq6}
\end{eqnarray}
where $\alpha = e^2/\hbar c$ is the QED fine structure 
constant,
$\lambdabar_C = \hbar/mc$ is the reduced Compton wavelength of an electron and 
\begin{equation}
E_{\rm crit} = {m^2c^3 \over e\hbar} = 1.6 \times 10^{16}\ \mbox{V/cm} 
= 3.3 \times 10^{13}\ \mbox{Gauss}
\label{eq6a}
\end{equation}
is the QED critical field strength, discussed in
sec.~3.2 below.

Thus, our na\"\i ve quantum theory (classical electromagnetism plus $\hbar$)
leads us to expect important departures from classical electromagnetism
for waves of wavelength much shorter than the
Compton wavelength of the electron, and for 
fields of strength larger than the QED critical field strength.

\subsection{The Radiation Reaction and Compton Scattering}  

Compton scattering \cite{Compton} 
was one of the earlier predictions of quantum theory and its observation had 
an important
historical role in widespread acceptance of photons as quanta of light.
Compton scattering is distinguished from Thomson scattering of classical
electromagnetism in that  wavelengths of the photons involved in Compton
scattering are not small
compared to the Compton wavelength of the electron, when
measured in the frame in which the electron is initially at rest.
Hence Compton scattering appears to be exactly
the kind of example discussed above in which the radiation reaction should
be important.


A description of a quantum scattering experiment relates the 
energy and momentum (plus relevant internal quantum numbers) of the initial
state to those of the final state without discussion of forces.  Yet, we
can identify various correspondences between the quantum and classical
descriptions.

In the case of Compton scattering, the initial photon corresponds to the
external force field on the electron, while the final photon corresponds to 
the radiated wave.
The quantum changes in  momentum (and energy) of the electron in the
scattering process can be said to
correspond to classical time integrals of force 
(and of ${\bf F} \cdot {\bf v}$).
Conservation of momentum (and energy) is described in the scattering
process by including  the back reaction of the final photon on the electron
as well as the direct reaction of the initial photon.  Thus, the quantum
description, which incorporates conservation of momentum  (and energy), can be 
said to include automatically the (time-integrated) effects of the radiation
reaction.

Compton scattering is an electromagnetic scattering 
process in which large changes in momentum (and energy) of the electron are
observed (in the frame in which the electron is initially at rest).  It can 
therefore be said to correspond to a situation in which
the radiation reaction is large, in agreement with the semiclassical 
inferences of secs.~2 and 3.

The correspondence between quantum conservation of energy and the classical
radiation reaction appears to involve only the second term, $-Ru^\mu/c$,  in 
expression (\ref{eq111}) for the radiation-reaction force.
Since the electron has constant (though different) initial and final velocities
in a scattering experiment, the ``acceleration energy'' $Q$ of eq.~(\ref{eq102})
is zero both before and after the scattering, and the equivalent of $\dot Q$ 
cannot be expected to appear in the quantum description 
(at ``tree level'', in the technical jargon) of Compton scattering.

Effects corresponding to the near-field ``acceleration energy'' can be said to 
occur in quantum electrodynamics in the case of so-called vertex corrections
and propagator (mass) corrections, in which a virtual photon is emitted and
absorbed by the same electron.  These ``loop corrections'' to the behavior of
a quantum point charge implement the equivalent of the self interaction of an 
extended charge, but diverge when the emission and absorption occur at the same 
spacetime point.  They are the source of the famous infinities of QED that are 
dealt with by ``renormalization''.  See also sec.~4.1 below.

In the early 1940's, Heitler \cite{Heitler41,Heitler} and coworkers
\cite{Wilson41,Power} formulated a version of QED in which
radiation damping played a prominent role.  Following the suggestion of
Oppenheimer \cite{Oppenheimer}, they hoped that this theory would provide
a general method of dealing with the divergences of QED.
By selecting a subset of ``loop
corrections'', they deduced an expression for Compton scattering that
corresponds to classical Thomson scattering plus classical radiation damping.
While this result is suggestive, it does not appear to be endorsed in detail by
subsequent treatments of ``renormalization'' in QED.

\subsection{The Critical Field}

The second term under the radical in eq.~(\ref{eq6}) may be less familiar.  
The concept of a
critical field in quantum mechanics began with Klein's paradox \cite{Klein}: 
an electron
that encounters an (electric) potential step appears to be reflected
with greater than unit probability in Dirac's theory.

Sauter \cite{Sauter} noted that this effect arises only when the potential
gradient is larger than the critical field, $m^2c^3/e\hbar$.  The resolution
of the paradox is due to Heisenberg and Euler \cite{Heisenberg}, who noted
that electrons and positrons can be spontaneously produced in critical
fields -- a very extreme form of the radiation reaction.  
The critical field has been discussed at a sophisticated level by
Schwinger \cite{Schwinger51} and by Brezin and Itzykson 
\cite{Brezin70}, among many others.

An electron that encounters an electromagnetic wave of  critical strength
produces not only Compton scattering of the wave photons but also
electron-positron pairs.  These effects have recently been observed 
in experiments in which the author participated \cite{Bula,Burke}.

There is speculation that critical magnetic fields exist at the surface of
neutron stars \cite{Ginzburg,Woltjer,Chanmugam}, and may be responsible
for some aspects of pulsar radiation.

Pomeranchuk \cite{Pomeranchuk} noted that the Earth's magnetic field appears
to have critical strength from the point of view of an electron of energy
$10^{19}$ eV, which energy is at the upper limit of observation of cosmic
rays.

The critical field arises in discussion of the radiation,
commonly called synchrotron radiation, of electrons
moving in circular orbits under the influence of a magnetic field $B$.
If an electron of laboratory energy
${\cal E} \gg mc^2$ moves in an orbit with angular velocity $\omega_0$, then the
characteristic frequency of the synchrotron radiation is
\begin{equation}
\omega \approx \gamma^3 \omega_0,
\label{eq8}
\end{equation}
where $\gamma = {\cal E}/mc^2$ is the Lorentz boost to the rest frame of the
electron.  For motion in a magnetic field $B$,
 the cyclotron frequency $\omega_0$ can be written
\begin{equation}
\hbar\omega_0 = {mc^2 B \over \gamma B_{\rm crit}},
\label{eq9}
\end{equation}
where $B_{\rm crit} = m^2c^3/e\hbar$.  Thus, the characteristic energy of
synchrotron-radiation photons (often called the critical energy) is
\begin{equation}
\hbar\omega \approx {\cal E} {\gamma B \over B_{\rm crit}}.
\label{eq10}
\end{equation}
Hence an electron radiates away roughly 100\% of its energy in a single
synchrotron-radiation photon if the magnetic field in the electron's rest
frame, $B^\star = \gamma B$, has critical field strength.  In this regime a 
classical theory of
synchrotron radiation is inadequate \cite{Schwinger54,Schiff}.

Critical electric fields can be created for short times in the collision of
nonrelativistic heavy ions, resulting in positron production
\cite{Kozhuharov,Tsertos,Greiner}.

As a final example of the inapplicability of classical electromagnetism in
strong fields, the performance of future high-energy electron-positron
colliders will be limited by the disruptive (quantum) effect of the critical 
fields 
experienced  by one bunch of charge as it passes through the oncoming bunch 
\cite{beamstrahlung}.

\section{Discussion}

In this paper I have followed the example of Landau in using the argument of
sec.~2 to suggest limitations to the concepts of classical electrodynamics.
However, this line of argument appears to have played no role
in the early development of quantum mechanics. Rather, the argument was
used in the 1930's to suggest that quantum electrodynamics
might have conceptual limitation when carried beyond the leading order
of approximation 
\cite{Oppenheimer,Heisenberg,Heitler33,Bethe,Williams,Uehling,Nordheim}.
The history of this era has been well reviewed in the recent book by
Schweber \cite{Schweber}.  

While the program of renormalization, of which 
Lorentz was an early advocate in the classical context \cite{Lorentz06},
appears to have been successful in
eliminating the formal divergences that were so troublesome in the 1930's,
quantum electrodynamics is still essentially untested for fields in excess
of the critical field strength (\ref{eq6a}) \cite{Burke}.  It still may be
the case that this realm contains new physical phenomena that will validate
the cautionary argument of Oppenheimer \cite{Oppenheimer}.

We close with three examples to stimulate additional discussion.  Two are 
from strong-field electrodynamics; while
not necessarily suggesting defects in the theory, they indicate that not all 
aspects of QED are integrated in the most familiar presentations.  The
third example considers the case of extraordinarily short wavelengths.

\subsection{The Mass Shift of an Accelerated Charge}

We can rewrite the
nonrelativistic expressions (\ref{eq1}-\ref{eq1b}) for the radiation reaction as
\begin{equation}
{d \over dt} \left( m{\bf v} - {2e^2 \dot{\bf v} \over 3c^3} \right) =
{\bf F}_{\rm ext},
\label{eq200}
\end{equation}
and the relativistic expressions (\ref{eq110}-\ref{eq112}) as
\begin{equation}
{d \over ds} \left( mc^2 u^\mu - {2e^2 \over 3} {du^\mu \over ds} \right) =
F^\mu_{\rm ext} - {Ru^\mu \over c}.
\label{eq201}
\end{equation}
These forms suggest the interpretation that the momentum, $m{\bf v}$, of a
moving charge is decreased by amount $2e^2 \dot{\bf v}/3c^2$ if that
charge is accelerating as well \cite{Fulton,Rohrlich61}.  

If we take $mc$ as the scale of the ordinary momentum, then the effect of
acceleration, $eE/m$, due to an electric field $E$ becomes large 
in eq.~(\ref{eq200}) only when $E \gsim m^2c^4/e^3 = e/r_0^2$,
\ie, when the electric field is large compared to the classical critical 
field found in sec.~2.

This interpretation has been seconded by Ritus \cite{Ritus} based on a
semiclassical analysis (classical electromagnetic field, quantum electron)
of the behavior of electrons in a strong, uniform electric field.  He finds that
the mass of an electron (= eigenvalue of the mass operator) obeys
\begin{equation}
m = m_0 \left( 1 - {\alpha E \over 2 E_{\rm crit}} 
+ {\cal O}(E^2/E^2_{\rm crit})
\right),
\label{eq202}
\end{equation}
and remarks on the relation between this result and the classical
interpretations (\ref{eq200}-\ref{eq201}).  The mass shift of an
accelerated charge becomes large when $E \gsim E_{\rm crit}/\alpha = e/r_0^2$,
as found above.

The physical meaning of Ritus' result remains somewhat unclear.  For example,
a mass shift of the form (\ref{eq202}) does not appear
in Ritus' treatment of Compton scattering in intense wave fields \cite{Nikishov}
(which treatment agrees with other works), although the effective mass 
(\ref{eq95}) does appear.

\subsection{Hawking-Unruh Radiation}

According to Hawking \cite{Hawking},
an observer outside a black hole experiences a bath of thermal
radiation of temperature
\begin{equation}
T = {\hbar g \over 2\pi ck}\ ,
\label{eq301}
\end{equation}
where $g$ is the local acceleration due to gravity
and $k$ is Boltzmann's constant.  In some manner, the background
gravitational field interacts with the quantum fluctuations of the
electromagnetic field with the result that energy can be transferred to the
observer as if he(she) were in an oven filled with black-body radiation.
Of course, the effect is strong only if the background field is strong.

An extreme example is that if the temperature is equivalent to 1 MeV or more,
virtual electron-positron pairs emerge from the vacuum into real particles.

As remarked by Unruh \cite{Unruh}, this phenomenon can be demonstrated
in the laboratory according to the principle of equivalence: an accelerated
observer in a gravity-free environment experiences the same physics (locally) as
an observer at rest in a gravitational field.  Therefore, an accelerated
observer (in zero gravity) should find him(her)self in a thermal bath of 
radiation characterized by temperature
\begin{equation}
T = {\hbar a^\star \over 2\pi ck}\ ,
\label{eq302}
\end{equation}
where $a^\star$ is the acceleration as measured in the observer's
instantaneous rest frame.


The Hawking-Unruh temperature finds
application in accelerator physics as the reason that electrons in
a storage ring do not reach 100\% polarization despite emitting polarized
synchrotron radiation \cite{Bell}.  Indeed, the various limiting 
features of performance of a storage ring that arise due to quantum
fluctuations of the synchrotron radiation can be understood quickly in
terms of eq.~(\ref{eq302}) \cite{McDonald87}.

Here we consider a more speculative example.
Suppose the observer is an electron accelerated by an electromagnetic field 
$E$.
Then, 
scattering of the electron off photons in the apparent thermal bath 
would be interpreted by a laboratory observer as an extra contribution to the
radiation rate of the accelerated charge \cite{McDonald85}.
The power of the extra radiation, which I call Unruh radiation, is given by
\begin{eqnarray}
{dU_{\rm Unruh} \over dt} & = & \mbox{(energy flux of thermal radiation)}
\nonumber \\
& & \times \mbox{(scattering cross section)}.
\label{eq303}
\end{eqnarray}
For the scattering cross section, we use the well-known result for Thomson
scattering,
$\sigma_{\rm Thomson} = 8\pi r_0^2/3$.  The energy density 
of thermal radiation is given by the usual expression of Planck:
\begin{equation}
{dU\over d\nu}={8\pi\over c^3}{h\nu^3\over e^{h\nu /kT}-1},
\label{eq304}
\end{equation}
where $\nu$ is the frequency.  The flux of the isotropic radiation on the
electron is just $c$ times the energy density.  Note that these relations hold
in the instantaneous rest frame of the electron.  Then
\begin{equation}
{dU_{\rm Unruh}\over dtd\nu}={8\pi\over c^2}{h\nu^3\over e^{h\nu /kT}-1}
{8\pi\over 3}r_0^2.
\label{eq305}
\end{equation}
On integrating over $\nu$ we find
\begin{equation}
{dU_{\rm Unruh}\over dt}={8\pi^3\hbar r_0^2 \over 45c^2}\left({kT\over\hbar}
\right)^4={\hbar r_0^2 a^{\star 4}\over 90\pi c^6},
\label{eq306}
\end{equation}
using the Hawking-Unruh relation (\ref{eq302}).  The presence of $\hbar$ in
eq.~(\ref{eq306}) reminds us that Unruh radiation is a quantum effect.

This equals the classical Larmor radiation rate, 
$dU/dt = 2e^2a^{\star 2}/3c^3$, when 
\begin{equation}
E^\star = \sqrt{60\pi \over \alpha} E_{\rm crit} \approx
{E_{\rm crit} \over \alpha},
\label{eq308}
\end{equation}
where $E_{\rm crit}$ is the QED critical field strength introduced in
eq.~(\ref{eq6a}).  In this case, the acceleration $a^\star = eE^\star/m$ is 
about $10^{31}$ Earth $g$'s.

The physical significance of Unruh radiation remains unclear.  
Sciama \cite{Sciama} has emphasized how the apparent temperature of an
accelerated observed should be interpreted in view of quantum fluctuations.
Unruh radiation is a
quantum correction to the classical radiation rate that grows large only in
situations where quantum fluctuations in the radiation rate 
become very significant.
This phenomenon should be contained in the standard theory of QED, but a direct 
demonstration of this is not yet available.  Likewise, the relation between
Unruh radiation and the mass shift of an accelerated charge, both of which
become prominent at fields of strength $E_{\rm crit}/\alpha$, is not yet
evident.

The existence of Unruh radiation provides an interesting comment on the
``perpetual problem'' of whether a uniformly accelerated charge emits
electromagnetic radiation \cite{Ginzburg69}; this issue has been discussed
briefly in sec.~2.5.  The interpretation of Unruh radiation as a measure of
the quantum fluctuations in the classical radiation implies that the
classical radiation exists.  It is noteworthy that while discussion of
radiation by an accelerated charge is perhaps most intricate classically in
case of uniform acceleration, the discussion of quantum fluctuations is
the most straightforward for uniform acceleration.  

In addition, Hawking-Unruh radiation 
helps clarify a residual puzzle in the discussion
of the equivalence between accelerated charges and charges in a
gravitational field.  Because of the difficulty in identifying an
unambiguous wave zone for uniformly accelerated motion of a charge (in a 
gravity-free region) and also in the case of a charge in a uniform 
gravitational field, there remains some doubt as to whether the `radiation'
deduced by classical arguments contains photons.  Thus,
on p.~573 of the article by Ginzburg \cite{Ginzburg69}
we read: ``neither a homogeneous gravitational field nor a uniformly
accelerated reference frame can actually ``generate'' free particles,
expecially photons''.  We now see that the quantum view is richer
than anticipated, and that Hawking-Unruh radiation provides at least a
partial understanding of particle emission in uniform acceleration or
gravitation.   Hence, we can regard the concerns of Bondi and Gold
\cite{Bondi}, Fulton and Rohrlich \cite{Fulton}, the DeWitt's \cite{DeWitt64}
 and Ginzburg \cite{Ginzburg69} on radiation and the equivalence principle
as precursors to the concept of Hawking radiation.

\subsection{Can a Photon Be a Black Hole?}

While quantum electrodynamics appears valid in all laboratory studies so far,
which have explored photons energies up to the TeV energy scale, will this
success continue at arbitrarily high energies (\ie, arbitrarily short
wavelengths)?

Consider a photon whose (reduced) wavelength $\lambdabar$
 is the so-called Planck length \cite{Wheeler57},
\begin{equation}
L_P = \sqrt{\hbar G \over c^3} \approx 10^{-33}\ \mbox{cm},
\label{eq501}
\end{equation}
where $G$ is Newton's gravitational constant.
The gravitational effect of such a photon is quite large.  A measure of
this is the ``equivalent mass'',
\begin{equation}
m_{\rm equiv} = {\hbar \omega \over c^2} = {\hbar \over c\lambdabar} =
{\hbar \over cL_P} .
\label{eq502}
\end{equation}
The Schwarzschild radius corresponding to this equivalent mass is
\begin{equation}
R = {2G m_{\rm equiv} \over c^2} = {2 \hbar G \omega \over c^3 L_P} =
2 L_P = 2 \lambdabar.
\label{eq503}
\end{equation}
A na\"\i ve interpretation of this result is that a photon is a black hole if
its wavelength is less than the Planck length.  Among the scattering
processes involving such a photon and a charged particle would be the case
in which the charged particle is devoured
by the photon, which would increase the energy of the latter, making its
wavelength shorter still.  

At very short wavelengths, electromagnetism and gravitation become
intertwined in a manner that requires new understanding.  The current best
candidate for the eventual theory that unifies the fundamental
interactions at short wavelengths is string theory.  Variants of the
preceding argument are often used to motivate the need for a new theory.


\section{Acknowledgements}
 
The author wishes to thank John Wheeler and Arthur Wightman for discussions
of the history of the radiation reaction, Bill Unruh for discussions of the
Hawking-Unruh effect, and Igor Klebanov, Larus Thorlacius
and Ed Witten for discussions of string theory.

\vspace{-0.25in}     

\end{document}